\documentclass[twocolumn,showpacs,preprintnumbers,amsmath,amssymb]{revtex4}
\usepackage{graphicx}% Include figure files
\usepackage{dcolumn}% Align table columns on decimal point
\usepackage{bm}% bold math
\begin{document}

\title{Influence of spin-orbit interaction on quantum cascade
transitions}
\author{Vadim M. Apalkov}
\affiliation{Department of Physics and Astronomy, Georgia
State University, Atlanta, Georgia 30303, USA and
Department of Physics and Astronomy,
University of Manitoba, Winnipeg, Canada R3T 2N2}
\author{Anjana Bagga and
Tapash Chakraborty\renewcommand{\thefootnote}{*}\footnote{
Electronic mail: tapash@physics.umanitoba.ca}}
\affiliation{Department of Physics and Astronomy,
University of Manitoba, Winnipeg, Canada R3T 2N2}

\date{\today}
\begin{abstract}
We have investigated the effect of spin-orbit (SO) coupling on
the emission spectra of a quantum cascade laser. In an externally
applied magnetic field parallel to the electron plane, the SO
coupling would result in a double-peak structure of the optical
spectra. This structure can be observed within some interval of
magnetic fields and only for diagonal optical transitions
when the SO coupling is different in different quantum wells.
\end{abstract}
\maketitle

The quantum cascade laser (QCL) is a coherent source of infrared
radiation and also an ingenious demonstration of quantum confinement
and tunneling in quantum well structures \cite{Expt_review,theory_review}.
These are specially designed superlattices of quantum wells. Optical
transitions between the subband levels of dimensional quantization
in the growth direction of QCL occur within the active region. In
these subbands, motion of electrons in the growth direction is frozen
and electron motion is two-dimensional. The electron states within
each subband are characterized by a two-dimensioanl momentum,
$\vec{k}$, and optical transitions between subbands are allowed
only between the states with the same momentum $\vec{k}$ and the
same spin projection. It is now well established that a strongly
{\it asymmetric} confinement potential results in a spin-orbit
(SO) coupling \cite{bychkov84}. Many novel effects that are entirely
due to the SO interaction have been proposed and some are observed
experimentally \cite{spintro,ohno,tc}. In this paper we analyze the
possible effects of SO coupling on the optical emission of the QCL.

Since the SO interaction couples the orbital motion and spin one
would expect that SO coupling should produce two types of optical
lines, corresponding to transitions between the same spin orientation
of the two subbands and between the different spin orientations.
However, for a weak enough disorder only one type of transition is
allowed. This is because, for the SO interaction $\alpha (\vec{k}
\times\vec\sigma ) \vec{n}$, where $\alpha$ is the SO coupling constant,
$\vec\sigma$ is the spin operator, and $\vec{n}$ is
the unit vector normal to the two-dimensional plane \cite{bychkov84},
the spin direction is correlated with the direction of momentum and
the spin states will be characterized by definite values of the
chirality, i.e. the spin projection on the direction perpendicular
to $\vec{k}$. For a weak disorder the optical transitions are allowed
only between the states with the same $\vec{k}$. Then the requirement
of spin conservation during optical transitions allows only
transitions between the states with the same chirality, i.e.
only a single optical line should be observed.

To observe the two optical lines we need to modify the energy
spectra of electrons in different subbands. One way of doing this
is by applying a parallel magnetic field. Since we are studying the
qualitative effects of SO coupling on the optical spectra of the
QCL we consider only two subbands in the active region of the QCL.
Electrons in these subbands will have different positions in the
growth direction of QCL. In other words, denoting the growth direction
as $z$-axis, we assume that $z_u=\langle z\rangle$ (the average value
of $z$ for the upper subband) is different from $z_l=\langle z\rangle$
(that of the lower subband). The values of $z_u$ and $z_l$ depends
on the structure of the QCL and on the applied voltage. We will
consider these quantities as parameters of the problem. We also
assume that electrons occupy only the higher subband, and they are
in quasi-equilibrium with temperature $T$ and electron density
$n_s$. The wavefunctions of electrons in the upper and lower subbands
will then have the form $\Psi_u(x,y,z)=\psi_u(x,y)\chi_u(z)$ and
$\Psi_l(x,y,z)=\psi_l(x,y)\chi_l(z)$. Optical transitions between
the upper and lower subbands will then determine the emission spectra
of the QCL. The intensity $I$ of these transitions is proportional
to $|\left\langle \chi_u \right|z\left|\chi_l \right\rangle\left\langle
\psi_u |\psi_l\right\rangle|^2$. Since the SO coupling should manifest
itself in the $(x,y)$-planar dynamics we shall study below only the
$(x,y)$ part of this expression.

\begin{figure}
\begin{center}\includegraphics[width=8.5cm]{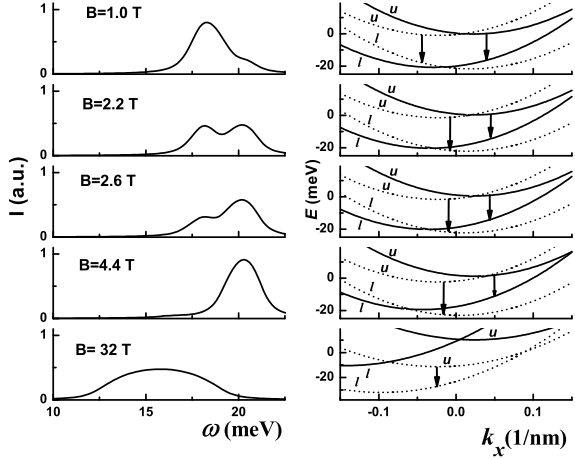}\end{center}
\vspace*{-1cm}
\caption{Emission spectra for different values of the parallel magnetic
field and for $\alpha_u=-\alpha_l=45$ meV.nm (left panel) and
the corresponding energy spectra of upper and lower subbands as a
function of $k_x$ for $k_y=0$ (right panel). States with a
positive value of $y$ projection of spin (solid lines),
and with a negative value (dotted line) are also shown.
The arrows illustrate two types of transitions which
results in two-peak structure of emission spectra. The letters
``u'' and ``l'' next to the lines stand for upper and lower
subbands, respectively.
}
\label{figone}
\end{figure}

To get a large SO coupling the quantum wells (QWs) in the active
region should be asymmetric. For such a structure the observed values
of the SO coupling constant lie in the range of 5 - 45 meV.nm \cite{SO}.
With an applied parallel magnetic field the Hamiltonian describing
the electron dynamics in the $x-y$ plane for upper and lower subbands
is \cite{bychkov84}
\begin{equation}
\label{H}
{\cal H}_s=\frac{1}{2m^*}\left(\vec p-\frac ec\vec A \right)^2
+\frac{\alpha_s}{\hbar}\left(\left[\vec p -\frac ec\vec A\right]
   \times{\vec\sigma}\right) \vec n - \frac12g\mu_B B\sigma_y,
\end{equation}
where the index $s=u,l$ stands for upper and lower subbands
respectively, $\vec\sigma=(\sigma_x, \sigma_y, \sigma_z)$ is the
vector of Pauli spin matrices, $\alpha_s$ is the SO coupling constant
for an electron in the $s$-th subband, and $m^*$ is the electron
effective mass. In Eq.~(\ref{H}) we assumed that the SO coupling is
different in different subbands, an important assumption since
only in this case we could get the well-resolved double-peak structure
of the optical spectra. Different values of $\alpha$ in different
subbands correspond to diagonal optical transition, i.e. the electrons
in upper and lower subbands are localized in different quantum wells.
Magnetic field in Eq.~(\ref{H}) is applied in the $-\hat{y}$ direction.
As a next step we introduce the gauge $\vec A=(-Bz,0,0)$ and replace
$z$ by its average value $z_s$ for the $s$-th subband. Then the
eigenfunctions of the Hamiltonian [Eq.~(\ref{H})] are classified according
to the chirality, $\kappa=\pm 1$, and form two branches of the spectrum
\begin{eqnarray}
E_{s,\kappa} (\vec k) &=& \frac{\hbar ^2}{2m^*} \left[ k_y^2 +
\left(k_x +\frac{z_s}{l_B^2}\right)^2 \right] \nonumber \\
&+& \kappa\alpha_s \sqrt{\left(k_x+\frac{z_s}{l_B^2}+
\frac12\frac{g\mu_B B}{\alpha_s}\right)^2+k_y^2},
\label{E}
\end{eqnarray}
where $l_B=(c\hbar/eB)^{\frac12}$ is the magnetic length. The
corresponding eigenfunctions are
\begin{equation}
\psi_{s,\kappa}(\vec k)=\frac1{\sqrt2}\left(
\begin{array}{c}
 1 \\
-i \kappa \exp \left(i\phi_{s,\vec k}\right)
\end{array}
\right) e^{ik_xx + ik_yy},
\label{psi}
\end{equation}
where the angle  $\phi_{s,k}$ is related to $\vec{k}$ as
\begin{equation}
\tan \phi_{s,\vec k}=\frac{k_y}
{k_x+{z_s}/{l_B^2}+{\frac12g\mu_B B}/{\alpha_s}}.
\end{equation}

Taking into account the spin conservation during the optical
transitions we can write the emission spectra as
\begin{eqnarray}
I(\omega) &=& I_0
\int \frac{d\vec k}{(2\pi)^2}
\sum_{\kappa_1\kappa_2} f\left[E_{u,\kappa_1}(\vec k)\right]
\nonumber \\
&\times&\left|1+\kappa_1\kappa_2 e^{i\left(\phi_{u,\vec k}
-\phi_{l,\vec k}\right)}\right|^2
 \nonumber \\
&\times&
\delta\left(E_{u,\kappa_1}(\vec k)-E_{l,\kappa_2 }(\vec k) -
\hbar\omega\right),
\label{I}
\end{eqnarray}
where $f(\epsilon)=1/[\exp(\epsilon-\mu_F)/k_BT+1]$ is the Fermi
distribution function for electrons in the upper subband with the
chemical potential $\mu_F$, which corresponds to electron density
$n_s$ and temperature $T$. It is easy to see that for zero parallel
magnetic field, $\phi_{u,\vec k}=\phi_{l,\vec k}$ and the
spin part in Eq.~(\ref{I}) is non-zero only for $\kappa_1=\kappa_2$.
In this case optical transitions are allowed only between the states
with the same chirality $\kappa$. This is also the case when the
magnetic field is large enough so that the Zeeman term in the
Hamiltonian becomes larger than the SO term. Transitions between
different subbands are allowed for intermediate values of the
magnetic field, although the main transitions still come from the
states with the same chirality. For a high density of electrons
on the upper subband these transitions should give only a single
line even in the presence of a parallel magnetic field. This
situation can be changed if the population of the upper subband is
low enough so that the electrons occupy states with the lowest
energy. In the momentum space these states correspond to a circle
with radius $\alpha_s m^*/\hbar^2$. It is easy to analyze this case
by fixing the value at $k_y=0$ and studying the spectra as a function
of $k_x$. Then the energy $E_{u,\kappa}(k_x)$ has two minima and
transitions from these minima can give rise to two peaks. The natural
requirement to resolve these peaks is that the width of the peaks
should be smaller than the separation between them. The maximum
separation between the peaks will occur when the SO coupling
constants $\alpha_s$ have different signs in the upper and lower
subbands.

\begin{figure}
\begin{center}\includegraphics[width=8.5cm]{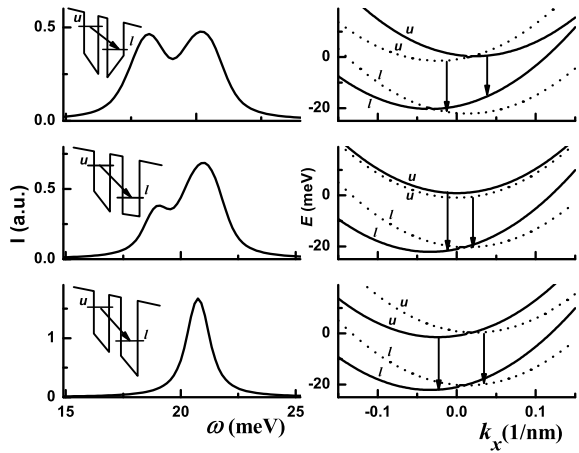}\end{center}
\vspace*{-1cm}
\caption{Emission spectra for different structure of active region of
QCL (left panel), which result in different values of SO coupling in upper
and lower subbands:
(a) $\alpha_l=-\alpha_u=45$ meV.nm,
(b) $ \alpha_u=0$ and $\alpha_l=45$ meV.nm,
(c) $\alpha_l=\alpha_u=45$ meV.nm. The schematic
illustration of corresponding active regions are shown as an inset.
The corresponding energy spectra of upper and lower
subbands as a function of $k_x$ for $k_y = 0$ are also shown (right panel).
States with a positive value of $y$ projection of spin (solid lines)
and those with negative value (dotted line) are also shown.
}
\label{figtwo}
\end{figure}

To analyze the possibility to observe SO-induced two-peak structure of the
emission spectra of a QCL we have calculated the optical spectra from
Eq.~(\ref{I}) for the smallest density of electrons on the upper subband
$n_s=10^{10}$cm$^{-2}$. To have the largest SO coupling constant, $\alpha\approx
45$  meV.nm, we assume that the QCL is based on the narrow gap semiconductor,
viz. InAs \cite{SO} ($m^*/m =0.042$ and $g = -14$). We have also fixed the
difference $|z_u-z_l|$ at 3 nm and study the optical spectra as a function of
the magnetic field. For illustration purpose we introduce the finite energy
difference between the energy levels (upper and lower subbands) of the size
quntization in the $z$ direction to be 20 meV. This means that without the SO
coupling and without a parallel magnetic field the emission spectra consists
of a single line centered at 20 meV.

In Fig.~1 the emission spectra are shown for $\alpha_u=-\alpha_l=45$ meV.nm
and for different values of the parallel magnetic field. In the right panel
the energy spectra of the upper and lower subbands are shown as a function of
$k_x$ for $k_y=0$. For $k_y=0$ the electron subbands can be classified by the
definite value of $y$-projection of the spin, $\sigma_y$. The solid lines
correspond to the positive value of spin, while the dotted lines correspond to
the negative values. Due to the small electron density in the upper subbands
only the lowest states are occupied. Because of the SO coupling electrons
in these states will have different directions of spin in different regions
of $k_x$. For example, for $k_x$ to the right from the the point of intersection
of two branches the spin is positive, while for $k_x$ to the left the
spin is negative. Transitions from these two types of electron states can
produce the two-peak emission spectra. These transitions are shown by arrows in
Fig.~1. At small values of the magnetic field these peaks almost coincide and a small
shoulder emerges due to allowed optical transition to the ground state. Eventually,
with increasing $B$ two peaks can be resolved and at $B\approx 2.2 T$ they have
the same intensity. At even larger $B$ the intensity of one of the peak
will be suppressed and the optical spectrum again acquires a single-peak
structure.

With increasing magnetic field the Zeeman energy becomes stronger
and only the states with negative spin is occupied. As a result,
there is only a single peak. This peak will be initially
blue shifted by an amount $\sim 5$ meV from the zero magnetic field
peak and then for a weak enough disorder it will
be red shifted as in the absence of any SO coupling \cite{apalkov01}.

\begin{figure}
\begin{center}\includegraphics[width=7.6cm]{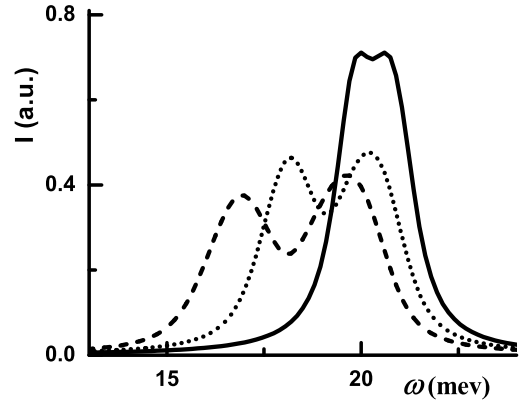}
\end{center}
\vspace*{-1cm}
\caption{Emission spectra for different values of $\alpha_u$ and
the magnetic field under the condition $\alpha_l=-\alpha_u$;
$\alpha_u=20$ meV.nm and $B=0.8$ tesla (solid line),
$\alpha_u=45$ meV.nm and $B=2.2$ tesla (dotted line), and
$\alpha _u = 60$ meV.nm and $B=3.24$ tesla (dashed line).
}
\label{figthree}
\end{figure}

The condition $\alpha_u=-\alpha_l$ results in the strongest separation
between two peaks. For smaller difference between $\alpha_u$ and $\alpha_l$
the two-peak structure becomes less pronounced and finally it will disappear
at $\alpha_u=\alpha_l$. The evolution of the two-peak emission spectra with
decreasing difference between $\alpha_u $ and $\alpha_l$ is shown in Fig.~2,
together with the energy spectra of upper and lower subbands. For $\alpha_u=0$
and $\alpha _l = 45 $ meV.nm the strongest effect that we can get at some value
of the magnetic field is the shoulder in the emission spectra [Fig.~2(b)].
For $\alpha_u=\alpha_l$, and for all values of the magnetic field there is
only a single peak [Fig.~2(c)]. While in this case there are also two types
of transitions, the width of the corresponding peaks are larger then the
separation between them. The inset in Fig.~2 illustrates schematically the
structure of two wells which gives the corresponding relation between the SO
coupling, where the upper and lower states are localized in different wells.

In Fig.~3 the evolution of a two-peak structure of the emission spectra
with change of SO coupling is shown for $\alpha_u=-\alpha_l$. The magnetic
field at which the two-peak structure becomes most pronounced is different
for different values of $\alpha_u$. With decreasing $\alpha_u$ the
separation between the peaks decreses and finally at small values of
$\alpha_u$ two peaks could not be resolved.

In conclusion, we have shown that the SO coupling in QCL could result
in a two-peak structure of the emission spectra. To observe such a structure,
the quantum wells constituting the active region of QCL should be asymmetric
and optical transitions should be diagonal. The next important condition is
that SO couplings in different quantum wells should be very different. In
this case a two-peak emission line can be developed within some interval of
a parallel magnetic field. Since in the  parallel magnetic field the
occupation of the upper subband determine the width of the optical lines,
to resolve the two-peak structure the electron density and the temperature
should be small enough, so that the width of the lines is less than the
separation between them.

The work of T.C. has been supported by the Canada Research Chair Program 
and the Canadian Foundation for Innovation Grant.

\end{document}